\documentclass[final]{aipproc}

\usepackage{amsmath,amssymb,graphicx,bm}

\newcommand{\be}{\begin{equation}}
\newcommand{\ee}{\end{equation}}
\newcommand{\lm}{\Lambda}
\newcommand{\kf}{k_{\rm F}}
\newcommand{\ef}{\varepsilon_{\rm F}}
\newcommand{\vlowk}{V_{{\rm low}\,k}}
\newcommand{\fm}{\, {\rm fm}}
\newcommand{\fmi}{\, {\rm fm}^{-1}}
\newcommand{\mev}{\, {\rm MeV}}

\layoutstyle{6x9}

\begin{document}

\title{Superfluidity in neutron stars and cold atoms}


\author{Achim Schwenk}{
address={TRIUMF, 4004 Wesbrook Mall, Vancouver, BC, Canada, V6T 2A3\\
Department of Physics, University of Washington, Seattle, WA 98195-1560}}

\begin{abstract}
We discuss superfluidity in neutron matter, with particular
attention to induced interactions and to universal properties 
accessible with cold atoms.
\end{abstract}

\maketitle

Superfluidity plays a central role in strongly-interacting 
many-body systems. Nuclear pairing shows striking trends 
in neutron-proton asymmetric systems~\cite{Litvinov}. The 
$\beta$ decay of the two-neutron halo in $^{11}$Li is suppressed 
due to pairing~\cite{Sarazin} similar to neutrino emission in 
neutron star cooling~\cite{Yakovlev}. Ultracold atoms exhibit 
vortices and superfluid characteristics in thermodynamic and 
spectroscopic properties~\cite{coldatoms}.

The physics of
dilute Fermi gases with large scattering lengths is universal,
independent of atomic or nuclear details. For neutrons the 
scattering length is also large, $a_{\rm nn} = - 18.5 \pm 0.3 
\fm$, and therefore cold atom experiments constrain 
low-density neutron matter. For instance, for two spin states 
with equal populations, the S-wave superfluid pairing gap of 
resonant gases of $^6$Li atoms, $^{40}$K atoms or neutrons is
given by $\Delta/\ef  = \zeta$, where $\ef = \kf^2/(2m)$ is 
the Fermi energy and $\zeta$ is a universal number.

\begin{figure}
\includegraphics[clip=,width=4.1in]{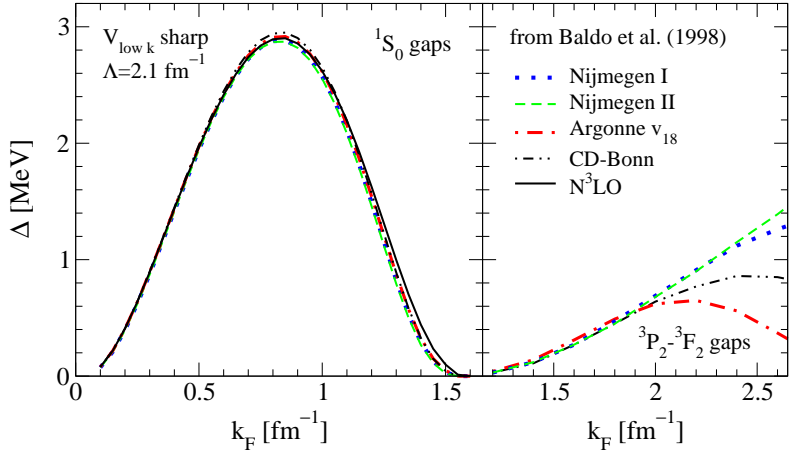}
\caption{The $^1$S$_0$ (left) and $^3$P$_2-^3$F$_2$ (right)
superfluid pairing gaps $\Delta \equiv \Delta(\kf)$ versus 
Fermi momentum $\kf$, based on various charge-dependent NN 
interactions at the BCS level. 
The results are for low-momentum interactions 
$\vlowk$ with $\lm=2.1 \fmi$~\cite{Vlowk1s0gap}
(left) or taken from Baldo {\it et al.}~\cite{Baldo} (right).}
\label{gaps}
\end{figure}

For relative momenta $k \lesssim 2 \fmi$, nucleon-nucleon (NN) 
interactions are well constrained by the existing scattering 
data~\cite{VlowkReport}. In Fig.~\ref{gaps}, we show
superfluid pairing gaps in neutron matter obtained 
by solving the BCS gap equation with a free spectrum. At low
densities (in the crust of neutron stars), neutrons form
a $^1$S$_0$ superfluid. At higher densities, the S-wave
interaction is repulsive and neutrons pair in the $^3$P$_2$
channel (with a small coupling to $^3$F$_2$ due to the tensor 
force). Fig.~\ref{gaps} demonstrates that the $^1$S$_0$ BCS
gap is practically independent of nuclear interactions, and
therefore strongly constrained by the NN phase shifts~\cite{Vlowk1s0gap}.
This includes a very weak cutoff dependence for the class of
low-momentum interactions $\vlowk$~\cite{VlowkReport} with 
sharp or sufficiently narrow smooth regulators with 
$\lm > 1.6 \fmi$. The model dependence for larger
momenta shows up prominently in Fig.~\ref{gaps} for the 
$^3$P$_2-^3$F$_2$ gaps at Fermi momenta $\kf > 2 
\fmi$~\cite{Baldo}.

Polarization effects (``induced interactions'') due to particle-hole
screening and vertex corrections are crucial for superfluidity. They
lead to a reduction of the S-wave gap, which is significant 
$[(4e)^{-1/3} \approx 0.45]$ even in 
the perturbative $\kf a$ limit~\cite{Gorkov}:
\be
\frac{\Delta}{\ef} = \frac{8}{e^2} \, \exp \biggr\{
\biggr( \, \begin{minipage}{4.4cm}
\includegraphics[scale=0.5,clip=]{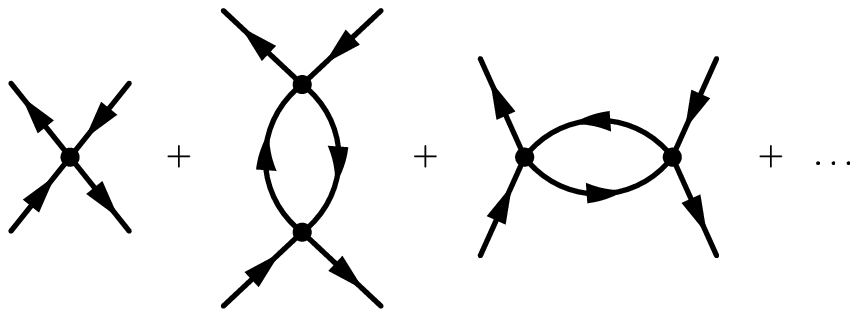}
\end{minipage}\biggr)^{-1} \biggr\}
= (4e)^{-1/3} \, \frac{8}{e^2} \, \exp \biggl\{ \frac{\pi}{2 \kf a} + 
{\mathcal O}(\kf a)\biggr\} \,. \nonumber
\ee
This reduction is due to spin fluctuations, which are repulsive for 
spin singlet pairing and overwhelm attractive density fluctuations.
In finite systems, the spin and density response differs. In nuclei 
with cores, the low-lying response is due to surface vibrations. 
Consequently, induced interactions may be attractive, since the
spin response is weaker.

The renormalization group (RG) provides a systematic tool to reduce 
a physical system to a simpler, equivalent problem focusing on 
relevant degrees of freedom. Following Shankar~\cite{Shankar},
we have applied the RG to neutron matter, restricting the effective 
interaction to low-lying states in the vicinity of the Fermi 
surface~\cite{RGnm}. Starting from the low-momentum interaction 
$\vlowk$~\cite{VlowkReport}, we solve a one-loop RG equation
in the particle-hole channels (``phRG'') that includes contributions 
from successive ph momentum shells. The RG builds 
up many-body effects similar to the two-body parquet equations, 
and efficiently includes induced interactions on superfluidity 
beyond the perturbative result.

\begin{figure}
\includegraphics[clip=,width=3.3in]{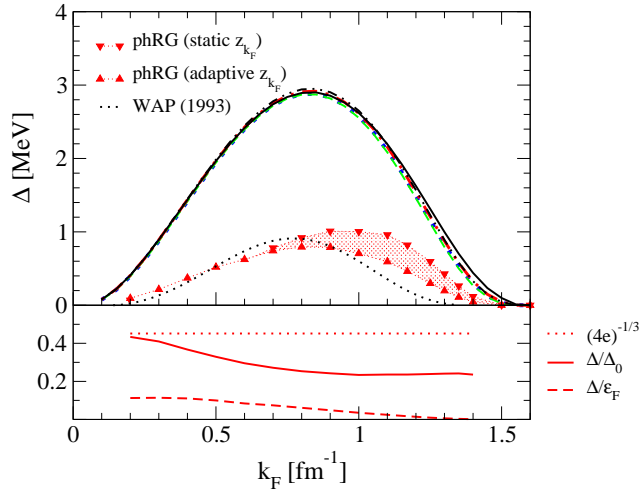}
\caption{Top panel: Comparison of the $^1$S$_0$ BCS gap to the
results including polarization effects through the phRG, for 
details see~\cite{RGnm}, and to the results of Wambach 
{\it et al.}~\cite{WAP}. Lower panel: Comparison of the full 
superfluid gap $\Delta$ to the BCS gap $\Delta_0$ and to the 
Fermi energy $\ef$.}
\label{indint}
\end{figure}

The phRG results for the $^1$S$_0$ gap
are shown in Fig.~\ref{indint}. We find a factor 
$3-4$ reduction to a maximal gap $\Delta \approx  0.8 \mev$. 
At the larger densities, the dotted 
band indicates the uncertainty due to an approximate self-energy 
treatment in~\cite{RGnm}. For the lowest densities, the phRG 
is consistent with the dilute result $\Delta/\Delta_0 
= (4e)^{-1/3}$. This is similar to the GFMC
calculations of Carlson {\it et al.}~\cite{GFMC} for cold 
atoms in the unitary regime, which are also consistent with 
the extrapolated dilute result to a good approximation.
On the lower side of Fig.~\ref{indint}, there are differences 
between neutron matter and unitary gases: For $\kf \approx 0. 4 \fmi$,
one has $\kf r_{\rm e} \approx 1$ (with  effective range $r_{\rm e}$),
and pairing is weaker, $\Delta/\ef \approx 0.1$. For these 
densities, neutron matter is close to the unitary regime, but 
theoretically simpler due to an appreciable effective range~\cite{dEFT}.
Note that the (low-order) CBF results of~\cite{AFDMC}
do not include long-range polarization effects, and
therefore are close to the BCS gap at low densities.

The RG approach is widely used in condensed matter physics to study 
the interference of different instabilities, especially in the 
context of the 2d Hubbard model. Similar competing instabilities
are present in color superconductivity at intermediate densities.
Here, the RG method seems ideal to resolve the zoo of possible phases.

Non-central spin-orbit and tensor interactions are crucial for
$^3$P$_2-^3$F$_2$ superfluidity. Without a spin-orbit interaction,
neutrons would form a $^3$P$_0$ superfluid instead.
The first perturbative calculation of non-central induced 
interactions shows that $^3$P$_2$ gaps below
$10 \, {\rm keV}$ are possible $({\rm while} \: \langle V_{\rm ind}
\rangle / \langle \vlowk \rangle < 0.5)$~\cite{3P2}. This
arises from a repulsive induced spin-orbit interaction due
to the mixing with the (large) spin-spin interaction.
Our result impacts the cooling of neutron stars~\cite{Yakovlev}
and would imply that core neutrons are only superfluid at late 
times $(t \sim 10^5 \, {\rm yrs})$.

\vspace*{2mm}

\noindent
I would like to thank B.~Friman, 
R.~Furnstahl, K.~Hebeler, C.~Horowitz, C.~Pethick and 
P.~Reuter for discussions and the organizers of 
QCHS VII for the stimulating meeting. This work is supported 
in part by NSERC and US DOE Grant DE--FG02--97ER41014. 
TRIUMF receives federal funding via a contribution agreement 
through the NRC.

\end{document}